\documentclass{aa}
\usepackage{graphics}

\begin{document}


\title{Statistical analysis of solar H$\alpha$ flares}

\author{M.~Temmer\inst{1}
 \and A.~Veronig\inst{1}
 \and A.~Hanslmeier\inst{1}
 \and W.~Otruba\inst{2}
 \and M.~Messerotti\inst{3}
}
\offprints{A. Veronig}
\institute{Institut f\"ur Geophysik, Astrophysik \&
 Meteorologie, Universit\"at Graz, Universit\"atsplatz 5, A-8010
 Graz, Austria
 \and Sonnenobservatorium Kanzelh\"ohe, A-9521
 Treffen, Austria
 \and Osservatorio Astronomico di Trieste, Via
 G. B. Tiepolo 11, I-34131 Trieste, Italy
 }
\date{Received 24 April 2001/ Accepted 12 June 2001}

\abstract{
A statistical analysis of a large data set of H$\alpha$ flares
comprising almost \mbox{100\,000} single events that occurred during the
period January 1975 to December 1999 is presented. We analyzed the flares evolution
steps, i.e. duration, rise times, decay times and event asymmetries.
Moreover, these parameters characterizing the temporal behavior of flares, as well
as the spatial distribution on the solar disk, i.e. \mbox{N-S} and \mbox{E-W}
asymmetries, are analyzed in terms of their dependency on the solar cycle.
The main results are: 1)~The duration,
rise and decay times increase with increasing importance class. The
increase is more pronounced for the decay times than for the rise
times. The same relation is valid with regard to the brightness
classes but in a weaker manner. 2)~The event asymmetry indices,
which characterize the proportion of the decay to the rise time
of an event, are predominantly positive ($\approx 90\%$). For about
50\% of the events the decay time is even more than 4 times as long as
the rise time. 3)~The event asymmetries increase with the importance
class. 4)~The flare duration and decay times vary in phase with the
solar cycle; the rise times do not. 5)~The event asymmetries do not
reveal a distinct correlation with the solar cycle. However, they
drop during times of solar minima, which can be explained by
the shorter decay times found during minimum activity.
6)~There exists a significant \mbox{N-S} asymmetry over
longer periods, and the dominance of one hemisphere over the
other can persist for more than one cycle. 7)~For certain cycles there
may be evidence that the N-S asymmetry evolves with
the solar cycle, but in general this is not the case. 8)~There exists
a slight but significant \mbox{E-W} asymmetry with a prolonged eastern excess.
}

\maketitle

\section{Introduction}

With the discovery of the first solar flare observed independently
by R.C. Carrington and R. Hodgson on September 1, 1859, knowledge
about these energetic phenomena on the Sun has steadily increased.
The statistical investigations of the characteristics of solar
H$\alpha$ flares essentially started in the 1930s, when a
worldwide surveillance of the Sun based on Hale's
spectrohelioscope was established (Cliver 1995). Since then many
papers have been published analyzing different statistical aspects
of solar flares. However, in most of the papers only single
aspects are investigated and/or the data set comprises a quite
limited period. Moreover, there exists no recent paper extensively
studying the statistical properties of solar H$\alpha$ flares.

In the present paper we make use of the substantial data
collection of solar H$\alpha$ flares in the Solar Geophysical Data
(SGD). We selected the period January~1975 (since then the
H$\alpha$ flares are listed with the same content and format) to
December~1999. With this selection we have a homogeneous data set
comprising almost \mbox{100\,000} single flare events, which
provides a significant statistical basis. Furthermore, as the
selected period entirely covers two solar cycles, 21 and 22, and
the rising phase of solar cycle~23 until the end of year 1999, the
data set also enables us to analyze dependency of the flare
characteristics on the solar cycle.

The paper is structured as follows. In Sect.~2 a characterization
of the data set is given. Sect.~3 describes the applied methods.
In Sect.~4 the results are presented and discussed, comprising a
statistical analysis of the temporal flare parameters combining
the data of the overall period, such as flare duration, rise
times, decay times (Sect.~4.1) and event asymmetries (Sect.~4.2).
Also, we analyze the above-mentioned temporal and spatial flare
characteristics, namely the N-S and E-W asymmetry, and its
dependency on the solar cycle (Sect.~4.3). Sect.~5 contains a
summary of the main results and the conclusions.
\section{Data}

For the present analysis we make use of the H$\alpha$ flares
listed in the SGD for January 1975 to December 1999. During this
period the occurrence of 97\,894 H$\alpha$ flare events is
reported. Table~\ref{impo_perc} lists the number of flares with
respect to the different importance classes, denoting the flare
size, i.e. subflares (S), flares of importance 1, 2, 3 and~4. In
Table~\ref{bright_perc} the number of flares is subdivided into
the three brightness classes, characterizing the intensity of the
flare emission in~H$\alpha$, i.e. f(aint), n(ormal) and b(right)
flares. Note that the total number of flare events subdivided into
importance classes and brightness classes differs from each other
as well as from the overall number of reported flares due to
incomplete early flare reports, where for some events the
importance class and/or the brightness class is not given.

\begin{table}
\centering
\caption{The number of flare events for the different
importance classes (S, 1, 2, 3, 4) and the corresponding
percentage values are listed. T denotes the total number of flares
occurring in the period 1975 to 1999, for which the importance
class is reported.}
\begin{tabular}{crr} \hline
Importance & No. of events &  No. (\%) \\ \hline
S          & 85649~~~~~ & 89.27~~  \\
1          & 9176~~~~~  & 9.56~~   \\
2          & 1014~~~~~  & 1.05~~   \\
3          & 101~~~~~   & 0.11~~   \\
4          & 5~~~~~     & 0.01~~   \\ \hline
T          & 95945~~~~~ & 100.00~~ \\ \hline
\end{tabular}
\label{impo_perc}
\end{table}

\begin{table}
\centering
\caption{The number of flare events for the different brightness classes
(f, n, b) and the corresponding percentage values are listed. T denotes the total
number of flares occurring in the period 1975 to 1999, for which the brightness class
is reported.}
\begin{tabular}{crr} \hline
Brightness & No. of events &  No. (\%) \\ \hline
f          & 59973~~~~~    &  61.42~~   \\
n          & 31454~~~~~    &  32.22~~   \\
b          & 6208~~~~~     &  6.36~~    \\ \hline
T          & 97635~~~~~    &  100.00~~  \\ \hline
\end{tabular}
\label{bright_perc}
\end{table}

It can be clearly seen in Table~\ref{impo_perc} that the
percentage of flares of importance 2, 3, and 4 is very small,
covering only $\approx\,$1.2\% of the overall number of events.
Especially during periods of minimum solar activity, their number
is vanishing. Therefore, to ensure a statistically meaningful data
set, for the analysis these three importance classes are merged
into one group, denoted as $>$1.

In Table~\ref{imp_bright} we list the number of flares belonging
to the different groups of importance and brightness classes,
resulting in nine subclasses. The table clearly reveals that the
importance and brightness classes show a strong interdependence.
Subflares have a strong tendency to be faint, importance~1 flares
to be of normal brightness and importance~$>$1 flares to be of
bright or normal brightness. Therefore, in the detailed analysis
we do not consider the partitioning of flares with respect to all
the subclasses but only with respect to the importance classes.
The choice of the importance instead of the brightness classes is
motivated by the fact that the importance classes are determined
in a more objective way, i.e. the measured area of a flare, than
the brightness classes, i.e. the subjective estimation of the
flare intensity by the observer.

\begin{table}
\centering
\caption{The number of flare events for the different brightness (f, n, b) and
importance (S, 1, $>1$) classes, given in absolute values and percentages of the
respective importance class.}
\begin{tabular}{lr@{~}rr@{~}rr@{~}r} \hline
         & \multicolumn{2}{c}{S} & \multicolumn{2}{c}{Imp. 1} &\multicolumn{2}{c}{Imp. $>$1} \\ \hline
f        & 56806 & (66.39\%) & 2251 & (24.69\%)  & 85  & (7.65\%) \\
n        & 25050 & (29.28\%) & 5098 & (55.93\%)  & 478 & (43.02\%) \\
b        & 3706  & (4.33\%)  & 1767 & (19.38\%)  & 548 & (49.33\%) \\ \hline
T        & 85562 & (100.00\%)& 9116 & (100.00\%) & 1111& (100.00\%) \\ \hline
\end{tabular}
\label{imp_bright}
\end{table}

The present statistical analysis deals with temporal as well as spatial
characteristics of H$\alpha$ flares. For the analysis of temporal flare
parameters, such as duration, rise times, decay times and event
asymmetries, we make use of letter codes listed in the SGD, which describe the
quality of the reported start, end and maximum times. The qualifiers
``U", ``E" and ``D" indicate that the given time is uncertain,
the event started before or ended after the reported time.
Additionally, digit qualifiers are annotated (from 1 to 9, given in
minutes), which describe the spread among the times reported by
different observatories. An asterisk denotes a spread of more
than 9 minutes.

For the analysis of temporal flare parameters events marked with an
``U", ``E", ``D" or an asterisk are discarded, in order to prevent biases due
to uncertainties in the reported times. Since the
data set is quite huge, this rejection of data has no distinct influence on the
statistics. Applying these selection criteria, we get 75\,739 H$\alpha$ flare events with
reported importance class, covering 68\,948 subflares (91.0\%), 6169 flares of
importance~1 (8.2\%) and 622 of importance~$>$1 (0.8\%). For the flares with reported
brightness class 76\,975 events are obtained. Note that for the analysis of spatial
flare parameters, such as \mbox{N-S} and \mbox{E-W} asymmetries, the original data set
as listed in Table~\ref{impo_perc} and Table~\ref{bright_perc} is used.

\section{Analysis and methods}

\subsection{Calculated parameters}

The H$\alpha$ flares that occurred within the considered time span
are statistically analyzed with respect to their temporal evolution.
From the evolution steps that are observed within a
flare event, i.e., start time~$t_{\rm start}$, end time~$t_{\rm end}$ and
maximum time~$t_{\rm max}$, we calculated the duration
($t_{\rm start} - t_{\rm end}$), the rise time ($t_{\rm max} - t_{\rm start}$)
and the decay time ($t_{\rm end} - t_{\rm max}$).
Furthermore, in order to characterize the proportion of the rise and
the decay time of a flare event, we computed the event asymmetry
index~$A_{\rm ev}$ (Pearce et al. 2001), defined as
\begin{equation}\label{Asymm}
 A_{\rm ev} = \frac{t_{\rm decay} - t_{\rm rise}}{t_{\rm decay} + t_{\rm rise}}\,,
 \label{ev_eq}
\end{equation}
with $t_{\rm rise}$ the rise time and $t_{\rm decay}$ the decay
time. The event asymmetry index is a dimension-less quantity
within the range $[-1,+1]$. A value close to zero states that the
rise and the decay time are roughly equal. The larger the
deviation from zero, the more asymmetric is the evolution of a
flare, whereas positive values indicate that the decay phase is
longer than the rising phase, for negative values, the opposite.

Furthermore, we analyzed the occurrence frequency of solar flares
in relation to the solar hemispheres. It is customary among authors
who are investigating \mbox{N-S} and \mbox{E-W} asymmetries to compute the
respective asymmetry indices (Letfus 1960).
The \mbox{N-S} asymmetry index~$A_{\rm NS}$ is defined as
\begin{equation}
A_{\rm NS} = \frac{n_{\rm N} - n_{\rm S}}{n_{\rm N} + n_{\rm S}} \,,
\label{ns_eq}
\end{equation}
with $n_{\rm N}$ and $n_{\rm S}$ the number of flares occurring in the
northern and southern hemisphere, respectively. The E-W asymmetry
index~$A_{\rm EW}$ is defined analogously. Asymmetry indices close to zero indicate
an equal distribution of flares with respect to the hemispheres. Positive values of
the \mbox{N-S} and the \mbox{E-W} asymmetry index characterize an excess of flares in
the northern and the eastern hemisphere, respectively.

\subsection{Statistical measures}

According to the distributions of the various temporal flare parameters, which are calculated
in the frame of this analysis, the determination of the average values was chosen
differently from the commonly used arithmetic mean. Since the relevant distributions
are significantly asymmetric (see Figs.~\ref{skewness}, \ref{flare_dur} and~\ref{asy_dist}), the
average is better represented by the median instead of the arithmetic
mean. Moreover, the median is more insensitive with respect to extreme data values than the
arithmetic mean. The median~$\tilde{x}$ of a distribution of values \{$x_{i}$\} is given
by that value, which has equal numbers of values above and below it. Note that
for distributions with positive skewness the median value is smaller than the
arithmetic mean, whereas for distributions with negative skewness it is larger.

Furthermore, as a robust measure of dispersion we applied the median
absolute deviation, which is more appropriate for the kind of distributions
we are dealing with than the standard deviation. The median absolute
deviation~$\tilde{D}$ is defined as
\begin{equation}\label{MAD}
 \tilde{D} = {\rm Median} \{|x_{i}-\tilde{x}|\} \, ,
\end{equation}
where \{$x_{i}$\} denote the data values and $\tilde{x}$ is the
median of the \{$x_{i}$\}.
Additionally, we also make use of the 95\% confidence interval,
which gives the probability that for a frequent use of the applied
procedure (which is legitimated by the huge data set), 95\% of the
data are gathered within the confidence limits. A conservative
95\%~confidence interval for the median is given by a rule of
thumb as $\tilde{x} \pm c_{95}$ with
\begin{equation}\label{conf}
c_{95} = \frac{1.58~(Q_{3}-Q_{1})}{\sqrt{n}} \, .
\end{equation}
$Q_{1}$ and $Q_{3}$ denote the first and the third quartile,
respectively, $n$ the total number of data values. The first
quartile $Q_{1}$, or 25th percentile, of a distribution is given
by that value which has 25\% of values below it; the third
quartile $Q_{3}$, or 75th percentile, is given by the value with
75\% of values below it. Note that the median~$\tilde{x}$ is
identical to the second quartile $Q_{2}$, or 50th percentile.

To specify the statistical significance of the N-S and E-W asymmetry indices,
we followed Letfus (1960). The dispersion of the N-S asymmetry of a random
distribution of flares is given by
\begin{equation}
\Delta A_{\rm NS} = \pm \frac{1}{\sqrt{2 (n_{\rm N} + n_{\rm S})}} \, ,
\label{dA_NS}
\end{equation}
which depends on the total number of flares occurring in the
northern and the southern hemisphere, respectively. To verify the
reliability of the observed \mbox{N-S} asymmetry values, a $\chi^{2}$ test is
applied with
\begin{equation}
\chi = \frac{2(n_{\rm N} - n_{\rm S})}{\sqrt{(n_{\rm N} + n_{\rm S})}}
     =  \frac{\sqrt{2} A_{\rm NS}}{\Delta A_{\rm NS}} \, .
\end{equation}
If $A_{\rm NS} = 2 \Delta A_{\rm NS}$, the probability that the observed \mbox{N-S}
asymmetry exceeds the dispersion value of a random distribution is $p=99.5\%$,
specifying asymmetry values which are highly significant. The same test, calculating
the corresponding quantities, is also applied to the \mbox{E-W} asymmetry indices.

\section{Results and Discussion}

\subsection{Duration, rise and decay time}

\begin{figure}
 \centering
 \vspace*{-0.3cm} 
    \resizebox{1.\hsize}{19cm}{\includegraphics{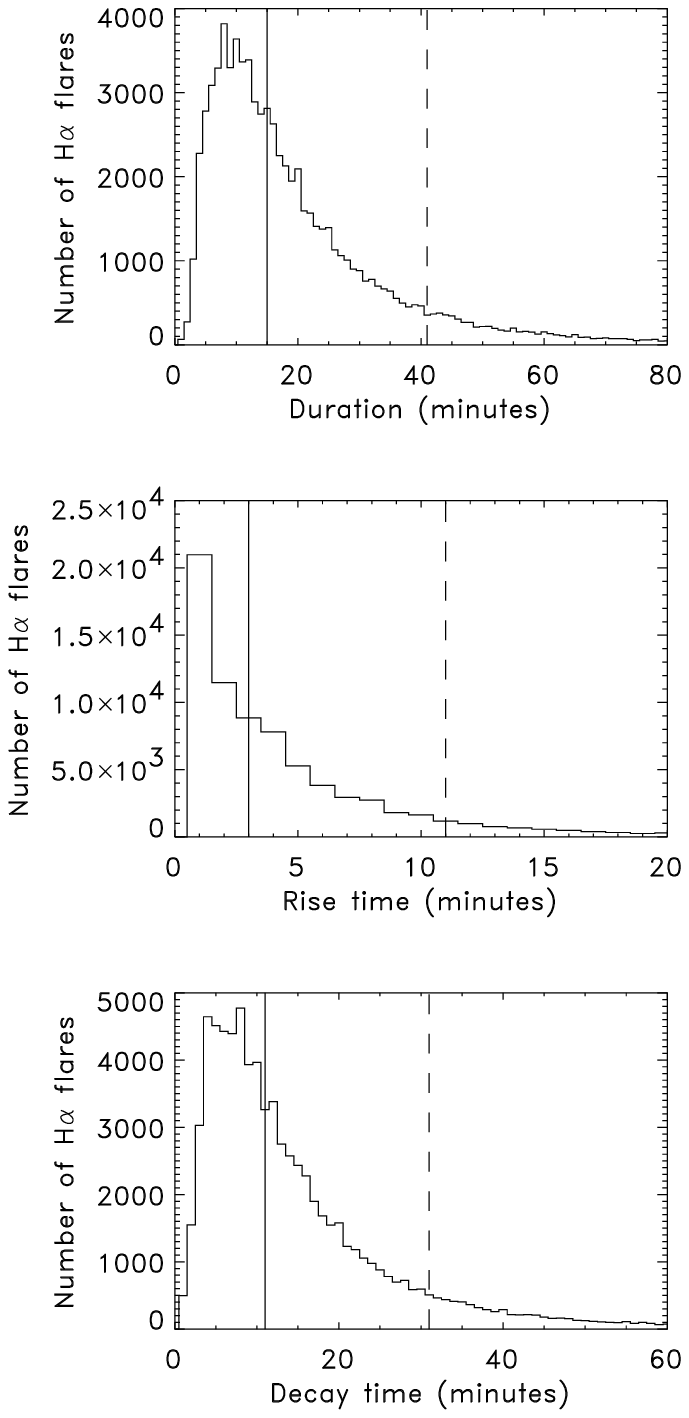}}
    \caption{Histograms of duration, rise and decay times calculated from
    the overall number of flares during 1975--1999 fulfilling the given
    quality criteria. The solid line indicates the median value
    of the distribution, the dashed line the 90th percentile. (Note, that for
    the sake of clearness, the ranges of the abscissa axes are cut off.)}
    \label{skewness}
\end{figure}

Fig.~\ref{skewness} shows the distributions of the duration, rise and decay
times calculated from all flares reported for the period 1975--1999, which
fulfilled the quality criteria regarding start, end and maximum time
stated in Sect.~2. All distributions reveal a pronounced positive skewness.
The overplotted solid line indicates the median value of the respective
distribution. The dashed line indicates the 90th percentile $P_{90}$, which
states that only 10\% of the events have a value larger than $P_{90}$.
In Table \ref{average} we give a list of various statistical measures characterizing
the duration, rise and decay times of the data set, namely the arithmetic mean, the median,
the mode (i.e. the most frequently occurring value), and the
90th percentile~$P_{90}$.

\begin{table}
\centering
\caption{Mean, median, mode and 90th percentile values of the duration, rise and
decay times of the total number of flares considered.}
\begin{tabular}{crrr} \hline
\multicolumn{1}{c}{Stat. measure} & \multicolumn{1}{c}{Duration} & \multicolumn{1}{c}{Rise time} & \multicolumn{1}{c}{Decay time}\\
               & \multicolumn{1}{c}{(min.)}   & \multicolumn{1}{c}{(min.)}    & \multicolumn{1}{c}{(min.)} \\ \hline
  Mean  & 20.6~~~~ & 5.1~~~~ & 15.5~~~~~ \\
  Median  & 15.0~~~~ & 3.0~~~~  & 11.0~~~~~ \\
  Mode  & 8.0~~~~  & 1.0~~~~  & 8.0~~~~~ \\
$P_{90}$& 55.0~~~~ & 11.0~~~~ & 31.0~~~~~ \\  \hline
\end{tabular}
\label{average}
\end{table}

\begin{figure}
 \centering
 \vspace*{-0.3cm}
   \resizebox{1.0\hsize}{19cm}{\includegraphics{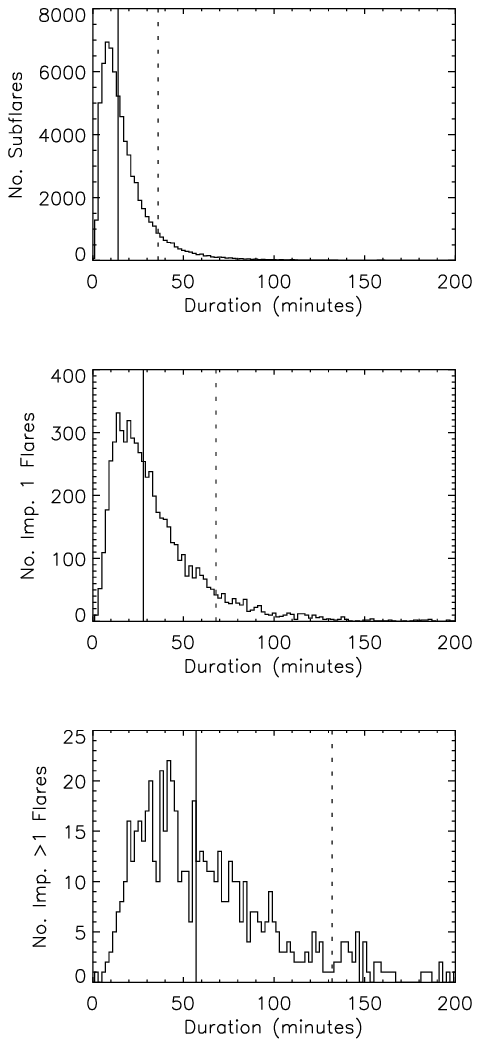}}
    \caption{Histograms of flare duration for the different importance classes (S, 1, $>1$).
    The solid line indicates the median value of the distribution, the dashed line
    the 90th percentile.}
    \label{flare_dur}
\end{figure}

Fig.~\ref{flare_dur} shows the distributions of the flare duration separately for
the different groups of importance classes, i.e., S, 1 and $>$1.
Table~\ref{impo_median(dur,rise,decay)} summarizes the
median values of the duration, rise and decay times with respect to the different
importance classes. Additionally, the 95\% confidence interval and the absolute median
deviation are given. For the sake of completeness, in
Table~\ref{bright_median(dur,rise,decay)} the same quantities are listed with regard
to the different brightness classes. As Table~\ref{impo_median(dur,rise,decay)} clearly
reveals, the duration, rise and decay times increase with the importance class. The same,
even if less pronounced, holds for the brightness classes, providing further
evidence that the importance and brightness classes are interdependent classification schemes.

\begin{table}
\centering
\caption{Median values with 95\% confidence interval, $~\tilde{x} \pm c_{95}$, and absolute median
deviation~$\tilde{D}$ of the duration, rise and decay times for the different importance
classes (S, 1, $>$1) and the total number of flares (T). All values are given in minutes.
(Note that due to the nonlinearity of the median the sum of the rise and decay time does not
exactly render the duration.)}
\begin{tabular}{cc@{$\,\pm\,$}c@{~\,}rc@{$\,\pm\,$}c@{~\,}rc@{$\,\pm\,$}c@{~\,}r} \hline
\multicolumn{1}{c}{Imp.} & \multicolumn{3}{c}{~Duration}  & \multicolumn{3}{c}{~Rise time} & \multicolumn{3}{c}{~Decay time} \\
     & ~~~$\tilde{x}$ & $c_{95}$& \multicolumn{1}{c}{$\tilde{D}$} & ~~$\tilde{x}$ & $c_{95}$& \multicolumn{1}{c}{$\tilde{D}$} & ~~~$\tilde{x}$ & $c_{95}$ & \multicolumn{1}{c}{$\tilde{D}$} \\ \hline
S    & 14.0 & 0.1 & 6.0  & 3.0 & 0.1 & 2.0 & 10.0 & 0.1 & 5.0 \\
1    & 28.0 & 0.5 & 12.0 & 5.0 & 0.1 & 3.0 & 22.0 & 0.5 & 10.0 \\
$>$1 & 57.0 & 3.2 & 24.0 & 8.0 & 0.6 & 4.0 & 45.0 & 2.9 & 21.0 \\ \hline
T    & 15.0 & 0.1 & 7.0  & 3.0 & 0.1 & 2.0 & 11.0 & 0.1 & 5.0  \\ \hline
\end{tabular}
\label{impo_median(dur,rise,decay)}
\end{table}

\begin{table}
\centering
\caption{Median values with 95\% confidence interval, $~\tilde{x} \pm c_{95}$, and absolute median
deviation~$\tilde{D}$ of the duration, rise and decay times for the different
brightness classes (f, n, b) and the total number of flares (T). All values are given in minutes.}
\begin{tabular}{cc@{$\,\pm\,$}c@{~\,}rc@{$\,\pm\,$}c@{~\,}rc@{$\,\pm\,$}c@{~\,}r} \hline
\multicolumn{1}{c}{Bright.} & \multicolumn{3}{c}{~Duration}  & \multicolumn{3}{c}{~Rise time} & \multicolumn{3}{c}{~Decay time} \\
& $~~~\tilde{x}$ & $c_{95}$& \multicolumn{1}{c}{$\tilde{D}$} & $~~\tilde{x}$ & $c_{95}$& \multicolumn{1}{c}{$\tilde{D}$} & $~~~\tilde{x}$ & $c_{95}$ & \multicolumn{1}{c}{$\tilde{D}$} \\ \hline
    f & 13.0 & 0.1 & 6.0  & 3.0 & 0.1 & 2.0 & 10.0 & 0.5 & 5.0  \\
    n & 19.0 & 0.2 & 9.0  & 4.0 & 0.2 & 2.0 & 14.0 & 0.2 & 7.0  \\
    b & 24.0 & 0.6 & 12.0 & 4.0 & 0.1 & 2.0 & 19.0 & 0.5 & 10.0 \\ \hline
    T & 15.0 & 0.1 & 7.0  & 3.0 & 0.1 & 2.0 & 11.0 & 0.1 & 5.0  \\ \hline
\end{tabular}
\label{bright_median(dur,rise,decay)}
\end{table}

The finding that the flare duration increases with increasing
importance class, hence with the flare area, is reported by a
number of papers (cf. Table~\ref{previous results}). Contrary to
that, by means of a scatter plot analysis of flare duration versus
flare area of an overall number of \mbox{16\,324} events, Yeung \&
Pearce (\cite{Yeung}) pointed out that there is a continuous
distribution of flare duration and flare area, concluding that
there is no evidence for more than one class of H$\alpha$ flares
and that there is no correlation between flare area and duration.
However, as can be seen in Fig.~\ref{flare_dur}, the distributions
of the flare duration reveal distinct differences for the
different classes. For subflares, the distribution reveals a
larger skewness and the maximum of the distribution is located at
smaller values than for larger flares. We suspect that the
continuous distribution of flare duration and flare area found by
Yeung \& Pearce (\cite{Yeung}) results from the fact that most of
the flares are subflares ($\approx 90\%$). Although the relative
number of subflares of long duration is very small (see
Fig.~\ref{flare_dur}, top panel), their absolute number is still
higher than those of large flares. (Due to the different scales of
the ordinate this fact does not show up in Fig.~\ref{flare_dur}).
Thus, if all flares are considered together, it is not possible to
separate the different distributions of different flare classes
and the analysis is mainly determined by the behavior of the
subflares, which intrinsically prevents detection of more than one
class of flare events.

\begin{table*}
{\centering
\caption{Mean values of flare duration calculated by different authors (given in minutes),
listed for the total number of flares (T) as well as separately for the importance classes
(S, 1, 2, 3, 4) if available. We list also the period of data record and the total number
of flares, on which the studies are based. Note that in particular for papers published before
1966, i.e. the year in which the current flare notation was constituted, importance
class~1 may also comprise subflares
(in former notation: $1-$) and class~3 may also comprise class~4 events (in former notation:
$3+$). In cases, in which it was possible to recover this fact, the concerned importance class
is transformed to the present notation.}
\begin{tabular}{lrrcccccc} \hline
  Author(s)  & \multicolumn{1}{c}{Period} & No. of flares & \multicolumn{6}{c}{Mean duration} \\
             &        &            &   S & 1 & 2 & 3 & 4 & T \\ \hline
 Newton \& Barton$^{1}$(\cite{Newbar})& 1935--1936 & ---  & ---  & ---   & ---  & --- & --- & 20.0--40.0 \\
 Waldmeier (\cite{Wald2})             & 1935--1937 &  357 & ---  & 21.0 & 38.0 & 61.0& --- & 27.0 \\
 Giovanelli (\cite{Giova})            & 1937--1938 &   24 & ---  &  ---  & ---  & --- & --- & 30.0 \\
 Waldmeier (\cite{Waldmeier})         & 1935--1944 &  927 & ---  & 20.3  & 33.4 & 62.4 & ---& 24.8 \\
 Ellison (\cite{Elli})                & 1935--1947 &  109 & ---  & 17.0  & 29.0 & 62.0&$\sim$180.0& ---\\
 Warwick (\cite{War})           & 1951--1953 &  357 & 31.0 & \multicolumn{4}{c}{60.0 ($\geq$1)} & 40.0 \\
 Dodson et al.$^{2}$ (\cite{Dodson})  & 1949--1952 &  194 & 28.0 & 43.0  & 66.0 & \multicolumn{2}{c}{84.0} & --- \\
 Waldmeier \& Bachmann (\cite{Bach})  & 1945--1954 & 1604 & ---  & 22.1  & 44.8 & 84.9& --- & --- \\
 Smith (\cite{Smith2})                & 1935--1954 & ---  & ---  & 20.3  & 33.4 & 62.4 & ---& --- \\
 Reid (\cite{Reid})                   & 1958--1965 & 2907 & 16.5 & 28.2  &\multicolumn{3}{c}{60.5 ($\geq$2)}& ---\\
 Ru\v{z}i\v{c}kov\'{a}-Topolov\'{a}$^{3}$(\cite{Ruzi})&1957--1965& 661   &---&---& 71.3 & 129.9& 305.5& ---\\
 Wilson (\cite{Wilson2})        & 1980       & 1348 & ---  & ---   & ---  & ---  &--- & 29.8\\
 Antalov\'{a} (\cite{Anti2})          & 1970--1974 &  460 & 27.0 & 55.0  & 78.0 & ---  & --- & --- \\
                                      & 1975--1979 &  561 & 30.0 & 51.0  & 78.0 & 201.0& ---& --- \\
 Wilson (\cite{Wilson})               & 1975       &  850 & 16.6 & 38.7  & 62.7 & ---  & --- & 18.1 \\
 Barlas \& Alta\c{s}$^{4}$ (\cite{Barlas})& 1947--1990& 3569&24.0& 38.0  &\multicolumn{3}{c}{78.0 ($\geq$2)}& ---\\
 Present paper                           & 1975--1999 & 75739 & 18.9&  35.7 & 66.3 & 116.0 & --- & 20.6 \\
\hline
\end{tabular}\\
}
 $^{1}$ The smaller values attribute to less intense, the larger values to more intense flares.\\
 $^{2}$ Duration calculated on the basis of photometric light curves. \\
 $^{3}$ Only great solar flares were considered. \\
 $^{4}$ Only spotless flares were considered. \\
\label{previous results}
\end{table*}

In Table~\ref{previous results} we give an overview of the studies
of flare duration that were carried out previously. Since in most
papers the arithmetic mean is used, we list in Table~\ref{previous results}
the mean values of the duration. For
the present analysis, we obtained the following mean values for
the flare duration: 18.9~min. for subflares, 35.7~min. for
importance~1, 66.3~min. for importance~2 and 116.0~min. for
importance~3 flares. For importance~4 flares we do not specify a
mean value as out of the four events of this class reported during
the considered period, only one event fulfilled the quality
criteria given in Sect.~2.

As can be seen in Table~\ref{previous results}, the duration
values obtained by the different authors reveal a quite large
dispersion. The reasons for this can be manifold. One possible
reason is given by observational selection effects which are
changing over the years. E.g., increasing time cadences in the
solar flare patrol lead to an increase in the number of detected
short-lived events, most of them subflares. Another aspect might
be the fact that in the miscellaneous studies, different data sets
are used, in which start, maximum and end times of a flare have
been derived/defined by different criteria. Moreover, it has to be
considered that the flare classification scheme has changed, from
(1, 2, 3) over ($1-$, 1, 2, 3, $3+$) to (S, 1, 2, 3, 4), which
sometimes leads to a mismatch in the comparison of values. (E.g.,
the old importance~1 class may comprise importance~1 flares as
well as subflares.) Finally, there may be also physical reasons,
i.e., the average flare characteristics, such as flare duration,
may change in the course of the solar cycle.

As we have a homogeneous data set at our disposal (for the data
collected in the SGD a constant classification scheme and a
well-defined methodology to derive start, end and maximum times is
in use), covering more than two solar cycles, we were able to
analyze the temporal flare parameters as a function of the solar
cycle (Sect.~4.3.2 and~4.3.3), with the objective to determine if
there are indeed physical changes in the temporal flare
characteristics.

\begin{figure}
 \vspace*{-0.4cm} 
 \centering
 \resizebox{\hsize}{18cm}{\includegraphics{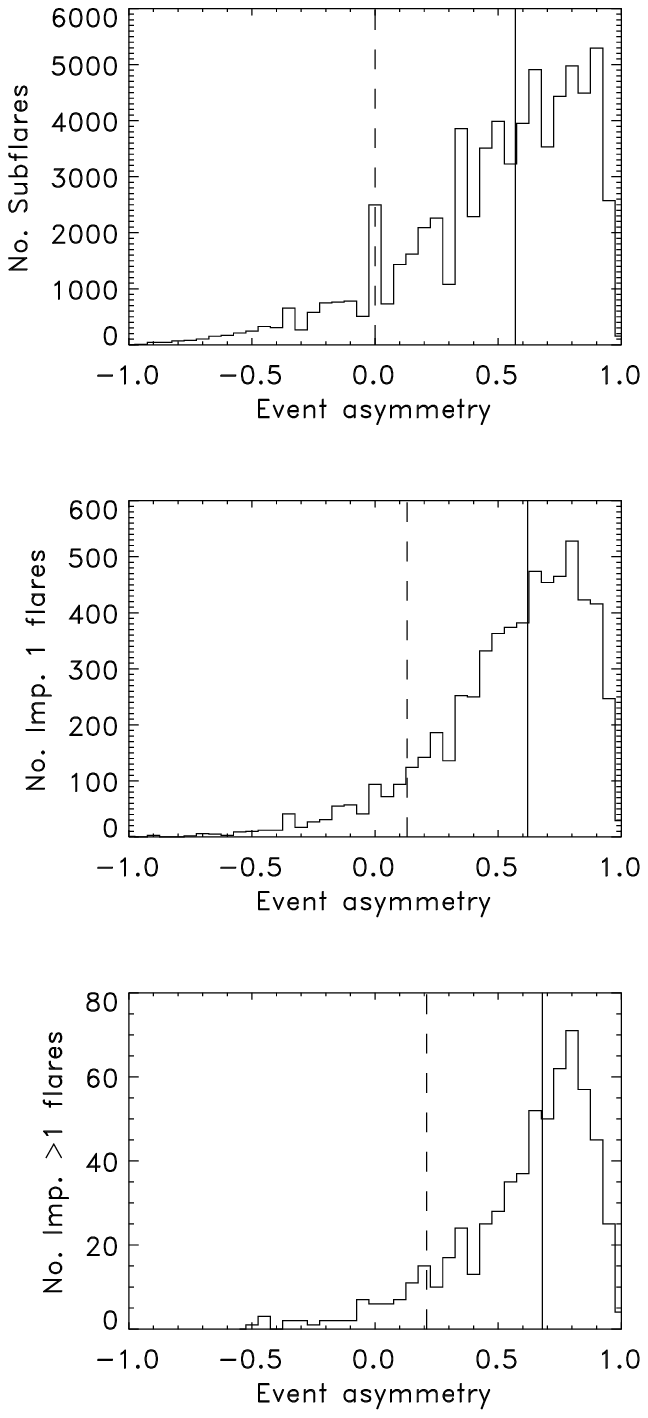}}
    \caption{Histograms of the event asymmetries, separately for
    importance classes~S, 1 and $>$1. The solid line indicates the
    median of the distribution, the dashed line the 10th percentile $P_{10}$,
    which indicates that only 10\% of the events have a value smaller than $P_{10}$.}
    \label{asy_dist}
\end{figure}

\subsection{Event asymmetry}

H$\alpha$ flares do not reveal a symmetrical behavior regarding
their temporal evolution, i.e. the rise time is not the same as
the decay time. With the use of the event asymmetry index defined
by Eq.~\ref{Asymm} we investigate this behavior quantitatively.
Fig.~\ref{asy_dist} shows the distributions of the event
asymmetries calculated for the different importance classes. All
distributions reveal a pronounced negative skewness, i.e. an
accumulation at positive values, showing that for the majority of
events the decay phase is significantly longer than the rising
phase.

\begin{table}
\centering
\caption{Median values of event asymmetries for the different
importance classes with 95\% confidence intervals. Furthermore,
the absolute median deviations and the 10th percentiles are
listed. For further discussions see the text.}
\begin{tabular}{cc@{$\,\pm\,$}c@{~~}c@{~~}c} \hline
Imp. & \multicolumn{4}{c}{~Event asymmetries} \\
     & ~~$\tilde{x}$ & $c_{95}$& $\tilde{D}$ & $P_{10}$ \\ \hline
  S  &  0.571 & 0.003 & 0.238 & 0.00 \\
  1  &  0.615 & 0.008 & 0.189 & 0.13 \\
$>$1 &  0.682 & 0.021 & 0.152 & 0.21 \\ \hline
 T   &  0.579 & 0.003 & 0.226 & 0.00 \\ \hline
\end{tabular}
\label{event_asymm}
\end{table}

In Table~\ref{event_asymm} we list the median values of the event
asymmetries for the different importance classes, as well as the
95\% confidence intervals, the absolute median deviations, and the
10th percentiles. It can be seen that the asymmetries increase
with increasing importance class. Since the differences of event
asymmetries between the various classes are larger than the
95\%~confidence intervals, the effect can be considered as
statistically significant. For the total number of events we
obtain a median event asymmetry of $\approx\,$0.6, which implies
that for about 50\% of the events the decay phase is more than
4~times as long as the rising phase. A value of $P_{10} \approx
0.0$ means that only about 10\% of all flares have a shorter decay
than rise time. Considering only flares of importance~$>$1, we
obtain for $P_{10} \approx 0.2$, stating that for about 90\% of
that type of events, the decay time is even more than 1.5 times
the rise time.

\subsection{Dependency on the solar cycle}
\label{dependency}

\subsubsection{Number of flares}

\begin{figure}
 \centering
 \resizebox{\hsize}{!}{\includegraphics{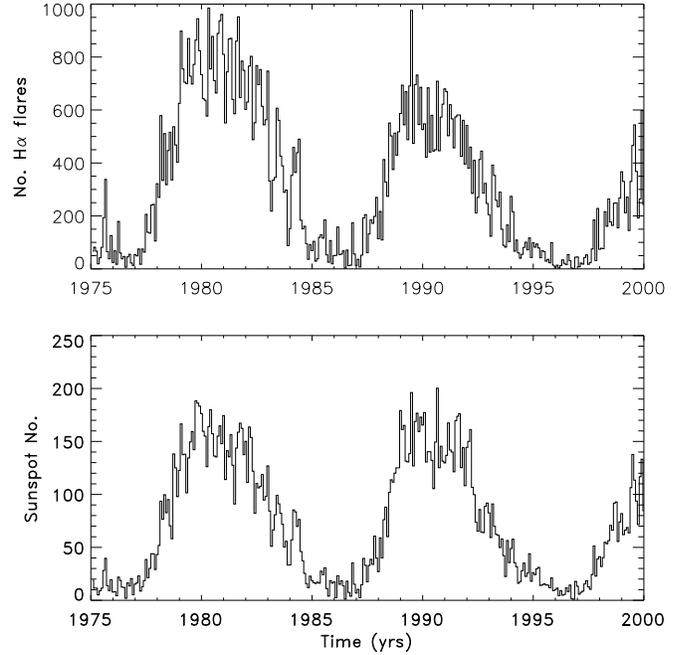}}
    \caption{Monthly number of H$\alpha$ flares (top panel) and
    monthly mean Sunspot Numbers (bottom panel) for the period January 1975 to
    December 1999. The data of the monthly mean Sunspot Numbers are taken from the SGD.}
    \label{SP_Halpha}
\end{figure}

\begin{figure}
 \centering
 \resizebox{\hsize}{!}{\includegraphics{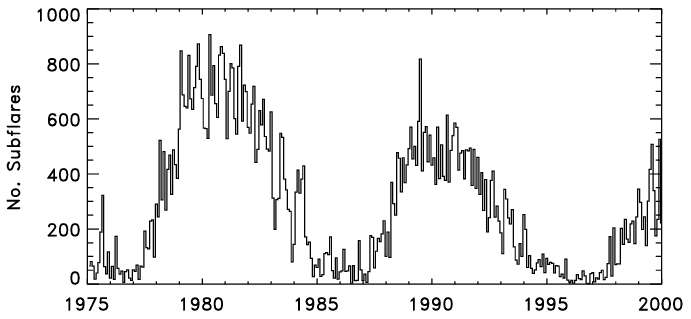}}
 \resizebox{\hsize}{!}{\includegraphics{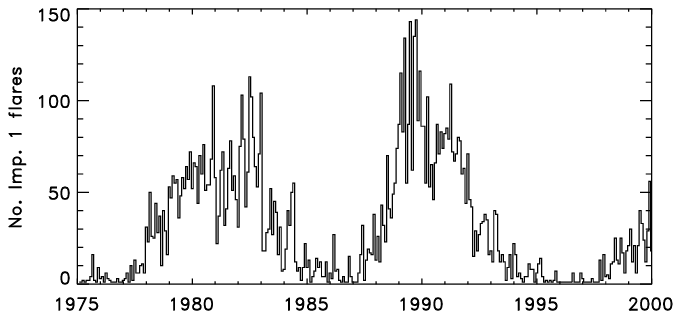}}
 \resizebox{\hsize}{!}{\includegraphics{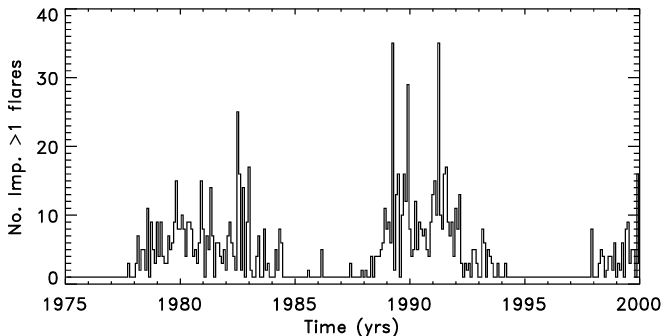}}
    \caption{Monthly number of H$\alpha$ flares, separately for subflares
    (top panel), flares of importance~1 (middle panel) and importance $>$1 (bottom panel).}
    \label{classes_month}
\end{figure}

In Fig.~\ref{SP_Halpha} we plot the monthly number of
H$\alpha$~flares and the monthly mean Sunspot Numbers for the
whole period of January 1975 to December 1999. The correlation
coefficient of the Sunspot Numbers with the total number of flares
occurring per month gives a value of $\approx\,$0.93, evidence
that the occurrence of H$\alpha$ flares is in strong coincidence
with the solar cycle.

Fig.~\ref{classes_month} shows the monthly number of flares
separately for the different importance classes. It is noteworthy
that, although cycles~21 and~22 do not differ remarkably regarding
the Sunspot Numbers, there is a significant difference in the
flare occurrence rate (see Figs.~\ref{SP_Halpha}
and~\ref{classes_month}). For the two solar cycles, which are
totally covered by the data, i.e. solar cycle~21 (June 1976 --
August 1986) and solar cycle~22 (August 1986 -- April 1996), we
list the average monthly rate of H$\alpha$ flares per cycle
(Table~\ref{flare_prod}). On the one hand, solar cycle~21 reveals
a conspicuously higher rate of subflares (which is also
responsible for the higher rate of the total of flares). On the
other hand, solar cycle~22 shows a higher rate of energetic
flares, i.e. events of importance~$\ge$1.

\begin{table}
\centering
\caption{Comparison of the average monthly flare rate for solar
cycles~21 and 22.}
\begin{tabular}{crr} \hline
Imp. & \multicolumn{1}{c}{cycle 21} & \multicolumn{1}{c}{cycle 22} \\ \hline
  S  & 388.6  & 259.2 \\
  1  &  34.6  & 37.2 \\
$>$1 &  4.2   & 4.9 \\ \hline
 T   &  427.4 & 301.3 \\ \hline
\end{tabular}
\label{flare_prod}
\end{table}

This fact that a cycle with a high rate of subflares produces
a smaller number of major flares than a cycle with less subflares
might be related to the mechanism of energy storage and
release during solar flares. In order to produce a large flare, an active region must
first store a vast amount of energy, which takes some time. The release of
this energy during a flare occurs either due to an internal instability
or an external destabilization, which can be caused, e.g., by nearby emerging
new magnetic flux or by a blast wave coming from another site of activity. Thus,
if there are many small releases of magnetic energy in the form of
subflares, there is little chance that enough energy can be stored
before the energy storage site is destabilized (\v{S}vestka 1995).

\subsubsection{Duration, rise and decay time}
\label{duration_y}

\begin{figure}
 \centering
 \resizebox{\hsize}{13.28cm}{\includegraphics{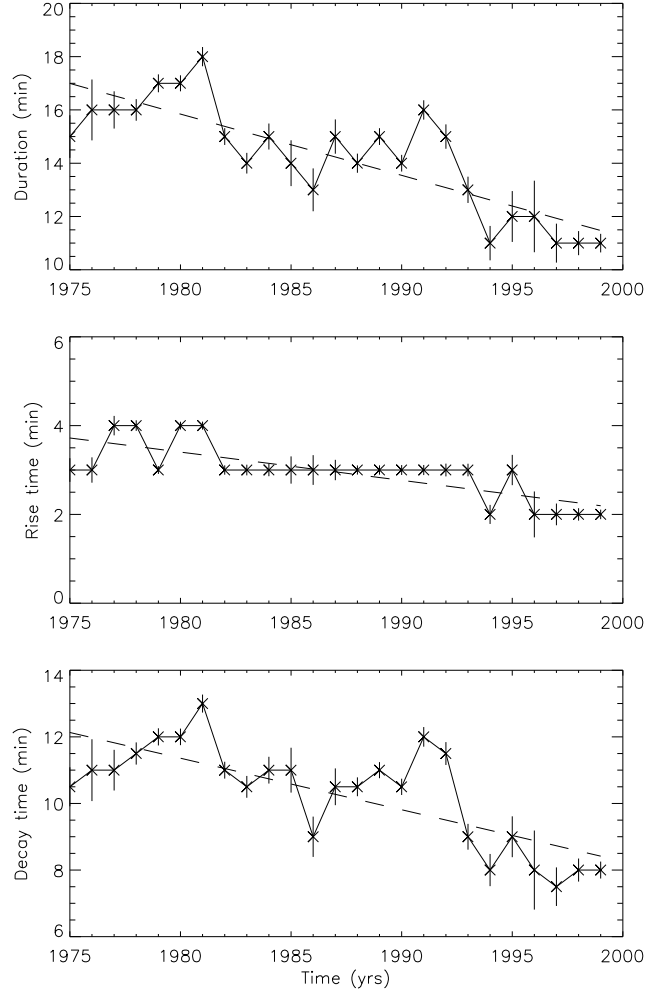}}
    \caption{Yearly median values of duration, rise and decay
    times. The error bars are given by the 95\% confidence
    intervals, the dashed lines indicate a linear trend.}
    \label{cycle_time}
\end{figure}

In Fig.~\ref{cycle_time} we plot the yearly median values of the
flare duration, rise and decay times from 1975 to 1999. It can be
clearly seen that each of the quantities reveals a trend to
decreasing values. We have plotted a linear least squares fit
superimposed on each of the time series in~Fig.~\ref{cycle_time}.
Obviously, such a trend does not represent a physical phenomenon
but we suggest that it is caused by observational selection
effects. In particular, the time cadences of solar flare
observations have increased with regard to former times, which
enables us to detect more short-lived events. Further evidence for
this explanation is provided by the comparison with a previous
paper of the authors (Temmer et al. 2000), in which the temporal
flare parameters of the subset 1994--1999 were investigated. For
the flares of importance 1 and $>$1, the calculated duration
values are very similar (the differences are of the order of the
95\% confidence intervals) but for the subflares, to which most of
the short-lived events belong, the average duration is distinctly
shorter ($\approx\,$10~min.) than those obtained by the present
analysis of the period 1975--1999 ($\approx\,$14~min., see
Table~\ref{impo_median(dur,rise,decay)}).

In order to find out if there is also a physical change in the
duration, rise and decay times with the solar cycle, we subtracted
the respective trends from the yearly duration, rise and decay
time values and calculated the correlation coefficients with the
yearly mean Sunspot Numbers (data taken from the SGD). The
correlation coefficients yield quite high values for the duration
as well as for the decay times, 0.71 and 0.73, respectively. For
the rise times it is distinctly lower, with a value of 0.40. This
means that the average duration and decay times are longer during
times of maximum activity than during minimum phases, whereas the
rise times do not show a noticeable variation with the solar cycle
(see also the time series in Fig.~\ref{cycle_time}). A similar
behavior is pointed out by Wilson (1987), who performed a
statistical study of flares occurring in the year 1980, i.e.
around solar maximum (Wilson 1982a, 1982b, 1983), and in the year
1975, i.e. around solar minimum (Wilson \cite{Wilson}). He found a
strong increase of the mean flare duration (1975: $18.1\pm
1.1$~min., 1980: $29.8\pm 2.2$~min.) and decay time (1975:
$12.9\pm 0.8$~min., 1980: $22.1\pm 1.7$~min.), and a slighter but
still pronounced increase of the mean rise time (1975:
\mbox{$5.2\pm 0.4$}~min., 1980: $7.7\pm 0.8$~min.) from solar
minimum to maximum.

We want to stress that the outcome of the present analysis that
the flare duration and decay times are significantly longer during
times of solar maximum than during solar minimum is not caused by
the fact that during maximum activity the number of large flares
with long duration is higher, as the number of large flares is too
low to significantly influence the overall statistics. However, to
be sure, we repeated the analysis considering only subflares,
which revealed the same behavior as the overall data set, i.e. a
significant change of the duration and decay times in the course
of the solar cycle.

\subsubsection{Event asymmetry}

\begin{figure}
 \centering
 \resizebox{\hsize}{!}{\includegraphics{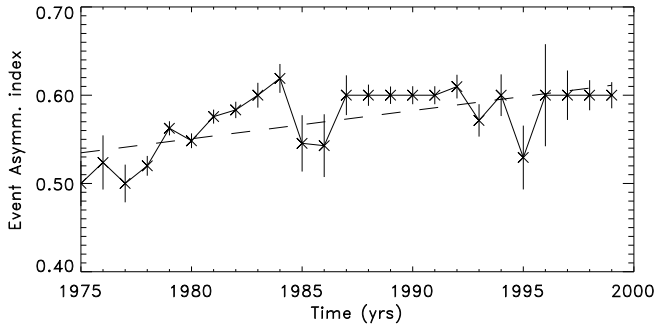}}
    \caption{Yearly median values of the event asymmetries. The error bars are given by the
    95\% confidence intervals, the dashed line indicates a linear trend.}
    \label{asy_yearly}
\end{figure}

In Fig.~\ref{asy_yearly} we plot the yearly median values of the
event asymmetries, defined by Eq.~\ref{ev_eq}. The time series
reveals a positive trend, resulting from the different trends of
the rise and decay times. Again, we calculated the correlation
coefficient with the yearly mean Sunspot Numbers after subtraction
of the trend, revealing no noticeable variation of the event
asymmetries coincident with the solar cycle.

However, as can be seen in Fig.~\ref{asy_yearly}, during minima of
solar activity the event asymmetries drop. This means that during
times of low solar activity on average the energy build-up and
decay of a flare takes place in a less asymmetric way than in
other phases of the solar cycle. In order to check if this
phenomenon is a statistical selection effect, induced by the low
number of large flares during minimum activity, which have larger
asymmetries than small ones (see Table~\ref{event_asymm}), we
repeated the analysis taking into account only subflares. The
behavior remained qualitatively the same, suggesting that the
decrease in the asymmetries during minimum activity is a real
effect. The drop can be explained by the fact that the decay times
are decreased during solar minimum phases but the rise times
basically remain constant during the solar cycle.

\subsubsection{North-South asymmetry}

\begin{figure}
 \centering
 \resizebox{\hsize}{!}{\includegraphics{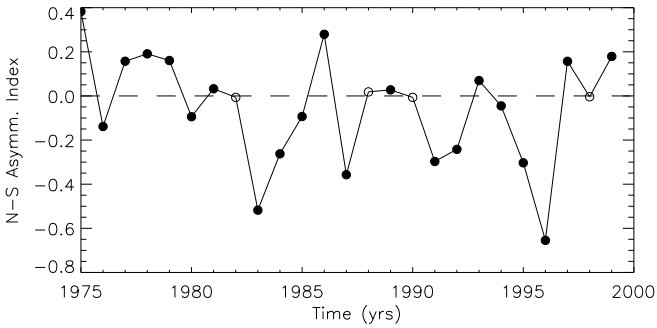}}
    \caption{Yearly \mbox{N-S} asymmetry index from 1975 to 1999. The significance of the
    \mbox{N-S} asymmetry values is specified by the plot symbols. Highly significant
    asymmetry values with $p \ge 99.5\%$ are marked with filled circles, otherwise
    white circles are drawn.}
    \label{asyindex(NS)}
\end{figure}

\begin{figure}
 \centering
    \resizebox{\hsize}{15.5cm}{\includegraphics{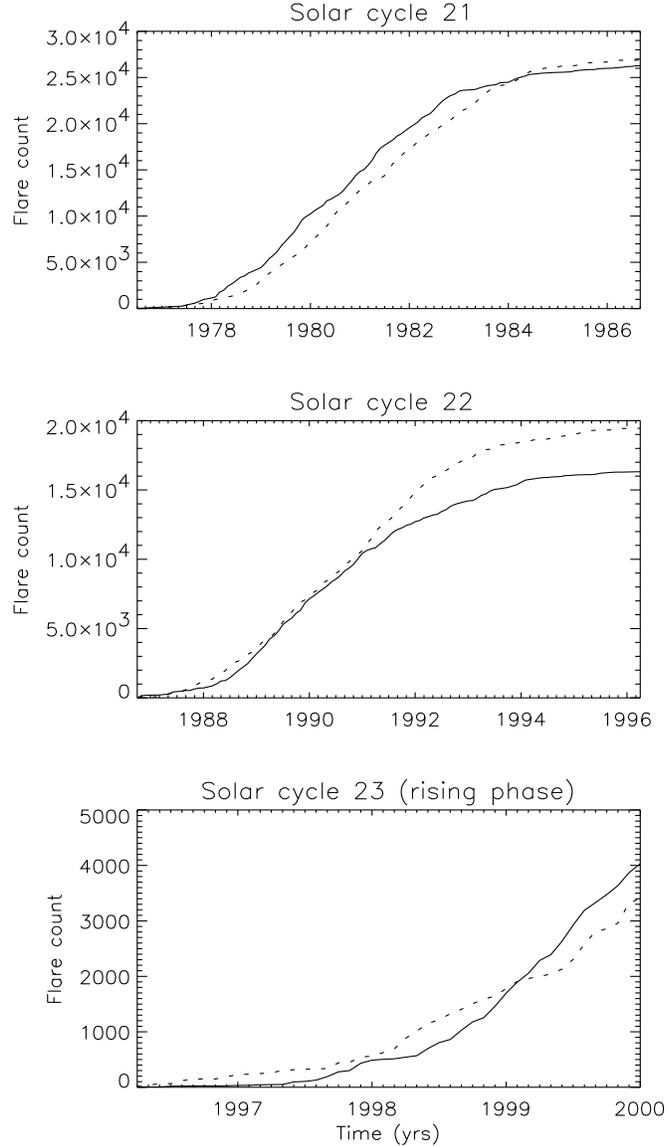}}
    \caption{Cumulative counts of flares occurring in the northern (solid lines)
    and the southern hemisphere (dashed lines) for solar cycles 21, 22 and 23.
    (For solar cycle~23 only the rising phase until the end of year 1999 is covered by the data.) }
    \label{ns_asymm}
\end{figure}

The existence of a \mbox{N-S} asymmetry of solar activity is
generally accepted even if the phenomenon is still not
satisfactorily interpreted. Several studies deal with the
\mbox{N-S} asymmetry of various kinds of manifestations of solar
activity, such as sunspots (Sunspot Numbers, areas, groups,
magnetic classes, etc.), flares, prominences, radio bursts, hard
X-ray bursts, gamma-ray bursts and coronal mass ejections.
Investigations of the \mbox{N-S} asymmetry of solar flares have
been carried out by  Ru\v{z}i\v{c}kov\'a-Topolov\'{a} (1974), Roy
(1977), Kno\v{s}ka (1984), Verma (1987), Garcia (1990),
Viktorinov\'a \& Antalov\'a (1991), Joshi (1995), Ata\c{c} \&
\"Ozg\"u\c{c} (1996), Li et al. (1998) and Ata\c{c} \&
\"Ozg\"u\c{c} (2001). All papers reveal the existence of a
\mbox{N-S} asymmetry; however, there are different outcomes if the
evolution of the \mbox{N-S} asymmetry is correlated with the solar
cycle or not.

In Fig.~\ref{asyindex(NS)} we plot the yearly \mbox{N-S} asymmetry
index, defined by Eq.~\ref{ns_eq}, for the years 1975--1999. 21
out of 25 asymmetry values turn out to be highly significant, with
a probability $p \ge 99.5\%$ that the observed asymmetry index
exceeds the dispersion value of a random distribution. The
correlation coefficient calculated from the \mbox{N-S} asymmetry
time series and the yearly mean Sunspot Numbers is very low,
showing that the \mbox{N-S} asymmetry does not evolve in
coincidence with the solar cycle.

In Fig.~\ref{ns_asymm} we give another kind of representation of
the \mbox{N-S} distribution, which was applied by Garcia (1990) to
soft X-ray flares. In the figure, the cumulative number of flares
occurring in the northern (solid lines) and the southern
hemisphere (dashed lines) during solar cycles~21 and~22 and the
rising phase of solar cycle~23 is plotted. The vertical spacing
between the two lines is a measure of the northern/southern excess
of H$\alpha$ flares up to that time.

As Fig.~\ref{ns_asymm} reveals, for solar cycle~21 an excess of flares
occurring in the northern hemisphere
develops during the rising phase of the cycle, staying roughly
constant during the phase of major activity. About three years before
the end of the cycle the northern excess degenerates and
finally a very slight southern excess remains (50.6\%~south, 49.4\%~north).
A same behavior but with a more pronounced spacing was found by
Garcia (1990) for soft X-ray flares~$\ge\,$M9 (see his Fig.~1), who
concluded from that result that the N-S asymmetry evolves in phase with
the solar cycle. Viktorinov\'a \& Antalov\'a (1991), who
analyzed soft X-ray LDE (long duration event) flares during solar cycle~21
also found that the larger number of flares in the northern hemisphere
is basically compensated by increased flare activity in the southern hemisphere in
the declining phase of the cycle.

However, for solar cycle~22 the situation is quite different, as
can be seen in Fig.~\ref{ns_asymm}. During the rising and maximum
phase of the cycle a slight southern excess is present, which is
strongly enhanced during the declining phase. Finally, for the
whole cycle a distinct southern excess remains (54.4\% south,
45.6\% north). A southern flare dominance during solar cycle~22
was also obtained by Joshi (1995) for H$\alpha$ flares and by Li
et al. (1998) for soft X-ray flares $\ge\,$M1. However, both
studies were constrained to the maximum phase of the cycle.

Considering the whole time period from the beginning of year 1975
to the end of year 1999, a change in the predominance of flare
occurrence in the northern and southern hemisphere can be
detected. During most of solar cycle~21, the northern hemisphere
was predominant. At the end of the cycle the dominance shifted to
the southern hemisphere, which prevailed also during the whole of
solar cycle~22. In the course of solar cycle~23 it seems that
after a strong peak of the N-S asymmetry indicating a pronounced
southern excess during the minimum phase 1996, the predominance
again shifts to the northern hemisphere (see
Figs.~\ref{asyindex(NS)} and~\ref{ns_asymm}). The preference for
the northern hemisphere during the rising phase of solar cycle 23
is also reported by Ata\c{c} \& \"Ozg\"u\c{c} (2001), who analyzed
the N-S asymmetry of the solar flare index. However, as no unique
relationship between the N-S asymmetry index as well as the
cumulative northern/southern flare counts and the solar cycle can
be found, the issue of whether the N-S asymmetry is in phase with
the solar cycle remains ambiguous.

\subsubsection{East-West asymmetry}

The existence of an \mbox{E-W} asymmetry of solar activity is a
more controversial issue than that of a N-S asymmetry. Since the
E-W distribution depends on the reference position, there is no
obvious physical reason why an \mbox{E-W} asymmetry should exist
for longer periods (Heras et al. 1990). However, several authors
(Letfus 1960; Letfus \& Ru\v{z}i\v{c}kov\'a-Topolov\'{a} 1980;
Kno\v{s}ka 1984; Joshi 1995) found evidence for the existence of a
small but significant \mbox{E-W} asymmetry in the occurrence of
H$\alpha$ flares. Heras et al. (1990) pointed out that the flare
events are not uniformly spread in longitude, stressing that this
fact has no influence on the \mbox{E-W} asymmetry but may be
attributed to the transit of active zones on the solar disk.
However, Heras et al. (1990) also found periods of pronounced and
prolonged \mbox{E-W} asymmetry, which exceed the expectations from
pure random fluctuations. Li et al. (1998), who studied the
\mbox{E-W} asymmetry of soft X-ray flares, found slight \mbox{E-W}
asymmetries, which they evaluated as not significant. However,
they pointed out that the flares are not uniformly distributed
over the visible longitude range.

\begin{figure}
 \centering
 \resizebox{\hsize}{!}{\includegraphics{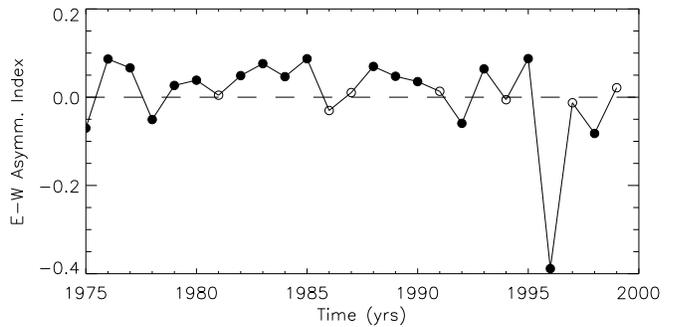}}
    \caption{Yearly \mbox{E-W} asymmetry index from 1975 to 1999. Highly significant
    asymmetry values with $p\ge 99.5$\% are indicated by filled circles.}
    \label{asyindex(EW)}
\end{figure}

In Fig.~\ref{asyindex(EW)} we plot the yearly \mbox{E-W} asymmetry
index, for the period 1975 to 1999. Even if the \mbox{E-W}
asymmetry values are rather low, 18 out of 25 asymmetry values
reveal a high statistical significance with $p \ge 99.5\%$. A
comparison with Fig.~\ref{asyindex(NS)} shows that the \mbox{E-W}
asymmetry is obviously lower than the \mbox{N-S} asymmetry.
However, the \mbox{E-W} asymmetry also can be quite pronounced
during certain periods, e.g., during the minimum phase in 1996 the
\mbox{E-W} asymmetry reveals a strong negative peak, indicating a
distinct western excess. At the same time, the N-S asymmetry
reveals a strong southern excess. An enhancement of the N-S and
\mbox{E-W} asymmetry during solar minima has already been reported
in several papers (Roy 1977, Kno\v{s}ka 1984, Heras et al. 1990,
Joshi 1995, Ata\c{c} \& \"Ozg\"u\c{c} 1996). However, we can
validate this effect only for the solar minimum in 1996. Moreover,
as the correlation coefficient of the \mbox{N-S} asymmetry time
series and the yearly mean Sunspot Numbers is very low, there is
no indication that in general the \mbox{E-W} asymmetry evolves in
phase with the solar cycle.

\begin{figure}
 \centering
    \resizebox{\hsize}{15.5cm}{\includegraphics{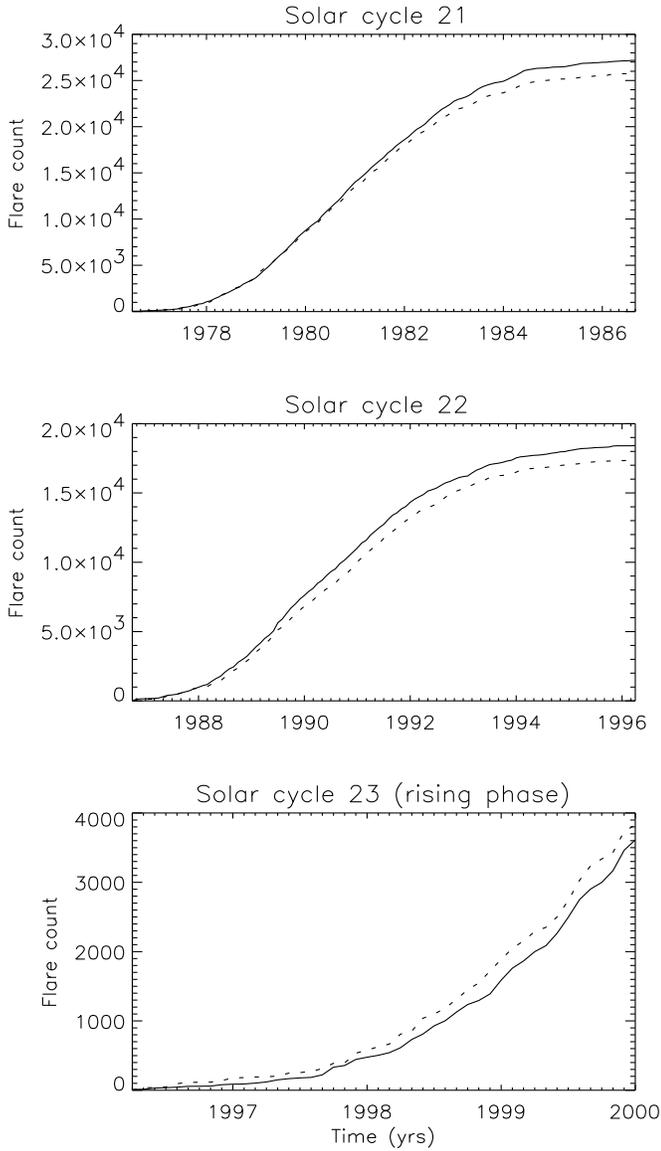}}
    \caption{Cumulative counts of flares occurring in the eastern (full lines)
    and the western hemisphere (dashed lines) for solar cycles 21, 22 and
    23 (until the end of year 1999).}
    \label{ew_asymm}
\end{figure}

In Fig.~\ref{ew_asymm} the cumulative number of flares occurring
in the eastern (solid lines) and the western hemisphere (dashed
lines) during solar cycles~21 and~22 and the rising phase of solar
cycle~23 is plotted. Solar cycle~21 and~22 reveal a rather similar
behavior. Except for the early phases of the cycles, a slight but
continuously increasing eastern excess develops. The final excess
amounts to 51.2\% east, 48.8\% west for solar cycle~21 and 51.5\%
east, 48.5\% west for cycle~22. For solar cycle~23, only a part of
the ascending phase is covered by the data. During this phase, a
western excess developed. However, this western excess may be
overruled by the following years of maximum activity, since the
number of flares at the end of 1999 covers just about
\mbox{15--20\%} of the overall flare number, which can be expected
for the whole cycle. Therefore, no definitive conclusion about the
E-W asymmetry of solar cycle~23 can be drawn.

Comprising previous investigations of the E-W asymmetry of
solar flares and the present paper, solar cycles 17 to 22 are
covered by the analysis (Letfus 1960: 1935--1958;
Ru\v{z}i\v{c}kov\'a-Topolov\'{a} 1974: 1957--1965; Letfus \&
Ru\v{z}i\v{c}kov\'a-Topolov\'{a} 1980: 1959--1976; Kno\v{s}ka
1984: 1937--1978; Heras et al: 1990: 1976--1985; Joshi 1995:
1989--1991; Li et al. 1998: 1987--1992; present paper:
1975--1999). Considering only those papers in which the found
\mbox{E-W} asymmetries are evaluated as significant, it is
worthwhile to mention that for longer periods, almost all authors
find an excess of flares in the eastern hemisphere. Thus, possibly
a low but prolonged dominance of the eastern hemisphere existed
during solar cycles~17 to~22, which is a rather surprising
outcome.

\section{Summary and Conclusions}

\label{discussion}

\subsection{Temporal parameters}

The main results of the analysis regarding temporal flare
parameters, as duration, rise times, decay times and event
asymmetries are summarized in the following. 1)~On average, the
duration, rise and decay times of flares increase with increasing
importance class. The increase is more pronounced for the decay
times (factor 4--5 between subflares and flares of
importance~$>$1) than for the rise times (factor 2--3). The same
relation holds for the brightness classes but in a weaker manner.
2)~The event asymmetries, which characterize the proportion of the
decay to the rise time of a flare, are predominantly positive ($
\approx 90\%$). For more than 50\% of all flares the decay phase
is even more than 4 times as long as the rising phase. 3)~On
average, the event asymmetries increase with the importance class.
4)~The duration changes in phase with the solar cycle, i.e. on
average the flare duration is longer during periods of maximum
activity than during solar minima. Since the rise times do not
reveal a distinct correlation with the solar cycle but the decay
times do, the variations of the duration in accordance with the
solar cycle are due to the variations of the decay times. 5)~The
event asymmetries do not reveal a distinct correlation with the
solar cycle. However, they decrease during solar minima.

The facts that on the average the flare duration increases with
the importance class and the rise times are shorter than the decay
times are reported in a number of previous papers (see the
references cited in Table~\ref{previous results}). However, the
new outcome of the present analysis is, on the one hand, that the
increase of the duration with the importance class in particular
results from the increase of the decay times, which is
significantly more pronounced than the increase of the rise times.
On the other hand, by the concept of the event asymmetry, we were
able to give a quantitative description of the asymmetry in the
flare development, which revealed that on average the asymmetry
also increases with the importance class. Both results suggest
that, with respect to the temporal behavior, the cooling phase of
the H$\alpha$ flare is more strongly affected by the flare size
than the phase of heating-up the chromospheric plasma at the flare
site.

Furthermore, a significant change in the duration and decay times
with the solar cycle was found. On average, during solar maximum
the decay times are larger than during solar minimum, with $t_{\rm
decay}^{\rm max} \approx 1.5 \cdot t_{\rm decay}^{\rm min}$. The
rise times do not reveal a significant variation in accordance
with the cycle. The combination of both facts can also account for
the drop in the event asymmetries found during solar minima. These
results suggest that the change in the flare duration is mainly
caused by the variations of the decay times, giving further
evidence that the flare cooling phase is more sensitive to changes
in the physical conditions of the chromospheric plasma than the
rising phase.

\subsection{Spatial distributions}

The main outcomes of the present analysis regarding the spatial
distribution of flares over the different hemispheres are:
1)~There exists a significant
\mbox{N-S} asymmetry over longer periods, and the dominance
of one hemisphere over the other can persist for more than one
cycle. 2)~For certain cycles there may be evidence that the N-S
asymmetry evolves in coincidence with the solar cycle, but in general this is
not the case. 3)~There exists a slight but significant \mbox{E-W}
asymmetry with a prolonged eastern excess during solar cycles~21 and~22.
Combining the results obtained by previous authors and the present paper,
possibly an eastern excess of solar flares existed during solar cycles~17 to~22.

The existence of a \mbox{N-S} asymmetry is generally accepted but still not
definitely interpreted. One possible explanation of the N-S asymmetry of solar activity
phenomena is that a time difference in the development of solar activity
on the northern and the southern hemisphere exists (e.g., Tritakis et al. 1997).
However, in this case the N-S asymmetry is expected to evolve in
coincidence with the solar cycle, which is rather ambiguous. Another explanation
utilizes the concept of ``superactive regions", which are large, complex, active
regions containing sunspots (Bai 1987, 1988). Such superactive regions
produce the majority of solar flare events and appear preferentially
in certain areas of the Sun, so-called ``active zones". As shown by Bai et al.
(1988) in the past, active zones were present on the Sun, which persisted
for several solar cycles. In that frame, the \mbox{N-S} asymmetry can be attributed
to the existence of active zones in the northern and southern hemispheres, which
can persist over long periods.

Also the fact that flares are not uniformly spread in heliographic
longitude over the solar disk (note that this non-uniformity has a
different meaning than the E-W asymmetry), pointed out by Heras et
al. (1990) and Li et al. (1998), can be understood in the
framework of the active zones concept. The longitudes, at which
the active zones are located, are more flare-productive than other
longitude ranges. The active zones induce a pronounced and
non-random flare activity, which is superimposed onto an episodic
and random flare activity coming from the other (less active)
regions of the solar disk (Heras et al. 1990).  However, the idea
of active zones cannot account for an \mbox{E-W} asymmetry
persisting over time scales larger than those of the solar
rotation.

The first report of an E-W asymmetry in solar activity was given
by Maunder (1907), who found that the total spot area and total
number of spot groups were larger in the eastern than in the
western hemisphere. A possible interpretation of this effect was
given by Minnaert (1946). A forward tilt of the vertical sunspot
axis causes a ``physical" foreshortening of the spots, which acts
more strongly on spots on the western than on those on the eastern
hemisphere. However, it cannot be seen how such an effect could
also account for an eastern excess in the flare occurrence rate.
In recent papers (Mavromichalaki et al. 1994, Tritakis et al.
1997) it has also been found that the emission of the eastern
hemisphere of the corona systematically predominates. However, the
coronal \mbox{E-W} asymmetry as well as the \mbox{E-W} asymmetry
of solar flares are still unexplained and controversial issues.

\acknowledgements

The authors thank Helen Coffey and Craig Clark from NGDC for
making available the H$\alpha$ flare data of the SGD.
M.T., A.V. and A.H. gratefully acknowledge the Austrian
{\em Fonds zur F\"orderung der wissenschaftlichen Forschung}
(FWF grant P13655-PHY) for supporting this project. M.M.
acknowledges the support by ASI and MURST.

\end{document}